\documentclass[
 amsmath,amssymb,
 aps,
prb,
]{revtex4-2}

\usepackage{amssymb}
\usepackage{dsfont}
\usepackage{epsfig}
\usepackage{amsmath}
\usepackage{subfigure}
\usepackage{graphicx}
\usepackage{textcomp}
\usepackage{url}
\usepackage{float}
\usepackage{color}
\usepackage{cases}
\usepackage[normalem]{ulem}

\usepackage[titletoc,title]{appendix}

\usepackage[colorlinks=true, urlcolor=blue, anchorcolor=blue, citecolor=blue,filecolor=blue,linkcolor=blue,menucolor=blue
]{hyperref}

\usepackage{color}

\def\Res{ {\rm Res}}

\def\<{\langle}
\def\>{\rangle}

\def\tr{{\rm   tr} }
\def\Im{{\rm Im}}
\def\Re{{\rm Re}}
\newcommand\encadremath[1]{\vbox{\hrule\hbox{\vrule\kern8pt
\vbox{\kern8pt \hbox{$\displaystyle #1$}\kern8pt}
\kern8pt\vrule}\hrule}} \def\enca#1{\vbox{\hrule\hbox{
\vrule\kern8pt\vbox{\kern8pt \hbox{$\displaystyle #1$} \kern8pt}
\kern8pt\vrule}\hrule}}
  \usepackage{bm}

\def\Xint#1{\mathchoice
   {\XXint\displaystyle\textstyle{#1}}%
   {\XXint\textstyle\scriptstyle{#1}}%
   {\XXint\scriptstyle\scriptscriptstyle{#1}}%
   {\XXint\scriptscriptstyle\scriptscriptstyle{#1}}%
   \!\int}
\def\XXint#1#2#3{{\setbox0=\hbox{$#1{#2#3}{\int}$}
     \vcenter{\hbox{$#2#3$}}\kern-.5\wd0}}

\def\dashint{\Xint-}

\usepackage{amssymb}

\usepackage{soul}
\usepackage{color}

\begin{document}

\title{A Riemann-Hilbert Approach to Slavnov Overlaps in the Lieb-Liniger model }

\author{Eldad Bettelheim }
\affiliation{Racah Inst.  of Physics, \\Edmund J. Safra Campus,
Hebrew University of Jerusalem,\\ Jerusalem, Israel 91904
}\date{\today}

\begin{abstract}
We provide a method to compute Slavnov overlaps in the Lieb-Liniger model using the steepest descent method of the Riemann-Hilbert problem. To do so, we employ the Matsuo-Kostov Representation of the Slavnov overlaps to write an integral equation for the respective resolvent, and then represent this  equation as a Riemann-Hilbert problem. We demonstrate the validity and applicability of the method by computing the Anderson orthogonality catastrophe in the $c\to\infty$ limit, corresponding to free fermions.
\end{abstract}

\maketitle

\section{Introduction}

Despite, in principle, requiring the ``fine-tuning'' of the parameters of the model such that problem may satisfy integrability, the Bethe ansatz has been shown to solve quite a large number of idealized physical systems. The list includes the Kondo model, the Lieb-Liniger model, the Hubbard model and many more (see, e.g., Ref. \cite{Korepin:Bogoliubov:Izergin:Quantum:Inverse:Scattergin:p264}).  The method is most efficient in finding thermodynamic potentials in equilibrium for such models. A thermodynamic  Bethe ansatz solution of an interacting many-body system is achieved when the a system of $N$ particles is reduced in an exact manner, in the thermodynamic limit, $N\to\infty,$ to a system of equation, where the computational  complexity involved  in solving those equations does not scale with $N$ at all. This is to be contrasted with a brute-force calculation where the computational complexity may be exponential in $N$.  

It is desirable to extend the Bethe-ansatz approach to non-equilibrium problems.  Additional objects appear in the analysis of non-equilibrium treatment, which go beyond the thermodynamic potentials. These objects are matrix elements and quantum overlaps. It is thus required to extend the Bethe ansatz method to allow to  reduce the computation of such objects in the thermodynamic limit to equations, the computational complexity of which not scaling with $N$, similar to the thermodynamic Bethe ansatz. Many strides have been made in this direction \cite{De:Nardis:Panfil:Form:Factors:Lieb:Liniger,Kormos:Mussardo:Trombettoni:Expectations:Lieb:Liniger,Kormos:Mussardo:Trombettoni:Lieb:Liniger:NonRelativistic,Koubek:Mussardo:Operator:Content:Sinh:Gordon,Caux:Calabrese:Slavnov:1P:Dynamical:in:LiebLinger,Kitanine:Kozlowski:Maillet:Slavnov:Terras,Kitanine:XXZ:Form:Factors,Dugave:Gohmann:Kozlowski:Thermal:Form:Factors,Smirnov:Book,Calabrese:Lieb:Liniger:Form:Factors} . 

The subject of this paper is to present a method which is quite general and may be applied to a number of integrable models. We present it here on the example of the Lieb-Liniger model, as to make the presentation as simple as possible. We treat, specifically Slavnov overlaps \cite{Slavnov}. Namely overlaps of Bethe states, where one of the states involved in the overlap is also assumed to satisfy the Bethe equations, namely is an eigen-state of an integrable Hamiltonian, while the other state does not. Many matrix elements and physically relevant overlaps can be represented in such a form \cite{Kojima:Korepin:Slavnov:Determinant:For:NonlinearSchroedinger,Calabrese:Lieb:Liniger:Form:Factors,Caux:Calabrese:Slavnov:1P:Dynamical:in:LiebLinger,Gorohovsky:Bettelheim:A:Derivation,Gorohovsky:Bettelheim:Announce,Gorohovsky:Bettelheim:Expectation:Values}. 

In order to find these overlaps we use a determinant representation due to Kostov and Matsuo \cite{Kostov:Inner:Product:Domain:Wall} which is distinct to the original determinant presented by Slavnov\cite{Slavnov}, but is more useful for our purposes. We are then tasked with computing the determinant in the thermodynamic limit. To this aim, we use the representation of the matrix involved as an operator acting on a certain function space, which was presented in Ref.\cite{Kostov:Bettelheim:SemiClassical:XXX}, to recast the problem as a problem of functional analysis. Then we present a novel Riemann-Hilbert representation of this problem. We then suggest to solve this problem using the steepest descent method of the Riemann-Hilbert problem due to Deift and Zhou  \cite{Deift:Its:Krasovksy:Toeplitz:Hankel,Deift:Its:Krasovsky:Toeplitz}. We show how this method works on the example of the Anderson orthogonality catastrophe. This amount to taking the coupling constant, $c$, of the Lieb-Liniger model to infinity to obtain a free fermion gas, and then taking the overlaps of two states, which differ in the      ``twist'', the phase, $e^{2\pi\imath\delta}$ ,  that the fermionic wave-function acquires when a particle completes a full cycle over the periodic one-dimensional system. The well known result, due to Anderson, is that the overlap should scale as $N^{-\delta^2}$\cite{Anderson:Catastrophe}. Although the example is presented in a certain limit, $c\to\infty$, the methods we use here to solve it are sufficiently representative of the method when $c$ is finite as well, while the benefit of this example is that we are thus able to compare to a well established result.

\section{Functional Approach to the Slavnov Determinant}

As mentioned in the introduction, we will be studying Slavnov overlaps in the Lieb-Linger model. We write $\tilde c$ for the coupling constant of the model. The Bethe equations {for the Bethe roots, $\theta_i,$ }in this model read:
\begin{align}
e^{\imath L\theta_j}\prod_{k=1}^{2N}\frac{\theta_j-\theta_k-\imath\tilde  c}{\theta_j-\theta_k+\imath\tilde  c}=-1\label{Bethe}
\end{align}
Assume a set of $2N$\ roots $\bm{\theta}$ satisfying the Bethe equations and another set of $2N$ roots $\tilde{\bm \theta}$ which do not necessarily satisfy the Bethe equations. To compute the overlap between those states, $\<\tilde{\bm{\theta}}|\bm{\theta}\>$, we use the Kostov-Matsuo formula\cite{Kostov:Inner:Product:Domain:Wall,Kostov:Bettelheim:SemiClassical:XXX}
\begin{align}
\<\tilde{\bm{\theta}}|\bm{\theta}\>=e^{\frac{\imath L}{2}\sum_j \theta_j-\tilde \theta_j}\det( \mathds{ 1} -  K),   
\end{align}
where the $4N\times 4N$ matrix $K$ depends only on the union, $\bm{\nu}$ of the two sets of roots ${\bm \nu}={\bm \theta}\cup { \tilde{\bm\theta}}$ and is given as follows:  
\begin{align} 
&K_{ij} = E_i\frac{1  }{\nu_{i} -\nu _j+ \imath\tilde  c }\label{Kdef}  \\
&E_i \equiv e^{-\imath L\nu_i} 
  {\prod_{k}(\nu_{i} -\nu_k+ \imath \tilde c)    \over \prod_{k\neq i}(\nu_{i}-\nu_k)}
\end{align}

We wish now to re-scale the complex plane in which the Bethe roots, $\nu_i$, live. To distinguish re-scaled variables we denote the latter in roman letters. Concretely, the re-scaling in question involves multiplying by the factor $\frac{L}{2\pi}$. Thus, for examples we associated the root $\nu_i$ the variable $u_i=\frac{ L\nu_i}{2\pi}$. We thus Obtain:
\begin{align}
&K_{ij} =\frac{E_i  }{u_{i} -u _j+ \imath\tilde  c }\label{Kdef}  \\
&E_i \equiv e^{-\imath 2\pi u_i} 
  {\prod_{k}(u_{i} -u_k+ \imath c)    \over \prod_{k\neq i}(u_{i}-u_k)}
\end{align}   
where $ c=\frac{L\tilde c}{2\pi}$

The matrix $K$\ can be given an operator form by the following trick (see Ref. \cite{Kostov:Bettelheim:SemiClassical:XXX}). We encode any vector $\bm{v}$ in $\mathds{C}^{2N}$ by a function $[\bm v](z)$ in the following way:
\begin{align}
[\bm v](z)=\sum\frac{v_i}{z-u_i}.
\end{align}
This simple mapping allows us to write $K$ in eq. (\ref{Kdef}) in a relatively simple functional form. To this aim first define $\varphi$ as follows:
\begin{align}
e^{\varphi(z)}=   e^{-\imath 2\pi z} 
\prod_l  {z-u_l+ \imath c   \over z-u_l}.\label{varphidef}
\end{align}This allows us to write the following functional representation of $K$:\begin{align}
[K\bm v](z)=\oint  \frac{e^{\varphi(z')}v(z'+\imath c)}{z-z'} \frac{dz'}{2\pi\imath},
\end{align}
the contour integral to be taken around any contour surrounding the $u_i$'s . To prove this formula, one needs only to consider the residues of the integrand at the poles, $u_i$. The Cauchy integral is necessary to remove the essential singularity at infinity introduced by $e^\varphi$.  

If we denote by $\bm e^{(j)}$ the vector with elements $e^{(j)}_i=\delta_{ij}$ then the function \begin{align}R(z,u_j)\equiv[(1-K)^{-1}\bm{e}^{(j)}](z)\label{Rdef}\end{align} solves:
\begin{align}
R(z,w)-\oint  \frac{e^{\varphi(z')}R(z'+\imath c,w)}{z-z'} \frac{dz'}{2\pi\imath}=\frac{1}{z-w},\label{ToSolve}
\end{align}
where $w$ should be set to $u_j$. The reason we write $w$ instead of $u_j$ is that this equation may be analytically continued (see Ref. \cite{Kostov:Bettelheim:SemiClassical:XXX}) to any $w$ in a neighborhood of the $u_i$  and so we may wish to solve Eq. (\ref{ToSolve}) treating $R$ as an analytic function of  $w$  in a neighborhood of the $u_i$'s.
The function\ $R(z,w)$\ is the functional representation of the resolvent, an object which  will feature heavily in the sequel.

The trace of the resolvent is essential in computing the determinant, by making use of the relation $d\log\det(1-K)=-\tr(1-K)^{-1}dK$.  The trace of the resolvent is easily computed in terms of an object we term $R_y,$ which is derived from  $R$\ defined in Eq. (\ref{Rdef}), by replacing $K$ by $e^{-Ny}K.$ Namely, we take: \begin{align}R_y(z,u_j)\equiv[(1-e^{-Ny}K)^{-1}\bm{e}^{(j)}](z).\label{Rydef}\end{align} This allows us to write\begin{align}
&\tr (\mathds{1}-e^{-Ny}K)^{-1}=\sum_j \underset{\lambda\to u_j}{\Res} R_y(z,u_j)=\oint\oint\frac{R_y(z,w)}{z-w}\frac{dz}{2\pi\imath}\frac{dw}{2\pi\imath} ,\label{trasint}
\end{align}
where the $w$ integral surrounds the $\lambda$ integral, which in turn surrounds the Bethe roots (see Ref.  \cite{Kostov:Bettelheim:SemiClassical:XXX} for a proof of this formula).
Now we can use (\ref{trasint}) to obtain:\begin{align}
&4N+\log\det(\mathds{1}-K)=4 N\int_0^{\infty} \tr (\mathds{1}-e^{-Ny}K)^{-1}dy\label{actualmaineq}=\\&\nonumber= 4N\int_0^{\infty}\oint\oint\frac{R_y(z,w)}{z-w}\frac{dz}{2\pi\imath}\frac{dw}{2\pi\imath} dy.
\end{align}

Another method to compute the determinant from the resolvent is to take a derivative to a translation of one set of the roots. Assume that $A$ is the set of indices of one of the two sets of roots, day $\tilde{\bm\theta}$. Namely, $\frac{L}{2\pi}\tilde{\bm \theta}=\{u_i\}_{i\in A}$. We can take a derivative which is associated with translating only this set of roots. We define this derivative as $\partial_\delta, $ such that we may write $
\partial_\delta=\sum_{i\in A}\partial _{u_i} 
$.

The derivative of $K$ with respect to $u_i$ may be written as:
\begin{align}
&\frac{1}{K_{ml}}\frac{\partial K_{ml}}{\partial u_i}=\delta_{mi}\sum_{n\neq i}\left(\frac{\delta_{n\neq k}}{u_i-u_n+\imath c}-\frac{1}{u_i-u_n}-2\pi\imath \right)+\delta_{m\neq i}\left(\frac{1}{u_m-u_i}-\frac{\delta_{l\neq i}}{u_m-u_i+\imath c}\right).
\end{align}
Define
 \begin{align}
 &\Lambda^{(i)}_{kl}=\delta_{kl}\delta_{k\neq i}\left(\frac{1}{u_k-u_i}-\frac{1}{u_k-u_i+\imath c}\right)+\delta_{ki}\left(-2\pi\imath+
\sum_{n\neq i}\frac{\delta_{n\neq i}}{u_i-u_n+\imath c}-\frac{\delta_{n\neq i}}{u_i-u_n}\right)\\
&v^{(i)}_{m}=\frac{1}{u_m-u_i+\imath c},
\end{align}
 so we can write
\begin{align}
& \partial_a \log\det(\mathds{1}-K)\nonumber=- \sum_{i\in A}\left[\tr (\mathds{1}-K)^{-1}\Lambda^{(i)} -\tr\Lambda^{(i)}+[(\mathds{1}-K)^{-1}v^{(i)}]_i\right].
\end{align} 

We shall make use of this equation below for the case of large $c$. In that case, we may neglect $v$ and all the terms that have $c$ in them, since $c$ is in the denominator, to obtain the following equations valid only approximately at large $c$:
\begin{align}
& \label{dlogdetda}\partial_\delta \log\det(\mathds{1}-K)+ 4 N= \sum_{i\in A,j\in A^C } \frac{R_{ii}+R_{jj}}{u_i-u_j}-\frac{2\delta_{i\neq j}}{u_i-u_j}.  
\end{align} 
We can write this as:
\begin{align}
\label{Fointment}\partial_\delta \log\det(\mathds{1}-K)+ 4 N +\sum_{i\in A,j\in A^C }\frac{2}{u_i-u_j}=\oint\oint\frac{F(z,w)}{w-z}\frac{dz}{2\pi\imath}\frac{dw}{2\pi\imath}
\end{align}
Where 
\begin{align}\label{OurF}
F(z,u_k)=\sum_{i\in A,j\in A^C} \frac{R_{ik}}{z-u_i}\frac{1}{u_j-u_i}-\sum_{i\in A^C,j\in A} \frac{R_{ik}}{z-u_i}\frac{1}{u_j-u_i},
\end{align}
 and $F(z,w)$ is obtained by analytical continuation of $F(z,u_k)$. Below we shall find $R(z,w)$ using the Riemann-Hilbert method and then deduce $F(z,w)$ from the knowledge of $R(z,w)$, for the particular example of the Anderson orthogonality catastrophe. 

\section{RH approach to matrix elements and expectation values}
The effectiveness of the method of representing the resolvent as an operator in a function space relies on being able to solve Eq.  (\ref{ToSolve}).
To achieve this aim, we shall first represent this equation as a Riemann-Hilbert problem. In fact, Eq. (\ref{ToSolve}) dictates the following equation for the jump discontinuity over the support of the Bethe roots as follows:
\begin{align}
R_+(z,w)=R_-(z,w)+\left(e^{\varphi_+(z)}-e^{\varphi_-(z)}\right)R_+(z+\imath c,w)
\end{align}
Based on this equation and given that $R$\ has only singularities on the real axis where the Bethe roots reside, we can write the following equation for the jump discontinuity of $R$ for $x$ real:
\begin{align}\begin{pmatrix}R(x-\imath0^+,w) &
\label{preRH}R_-(x+\imath( c-0^+),w) \\
\end{pmatrix}=
\begin{pmatrix}
R(x+\imath( c+0^+),w) & R(x+\imath 0^+,w)  \\
\end{pmatrix}\begin{pmatrix}e^{\varphi_-(x)}-e^{\varphi_+(x)} & 1\\1 & 0 \end{pmatrix}.
\end{align}

\begin{figure}[b]
\includegraphics[width=5cm]{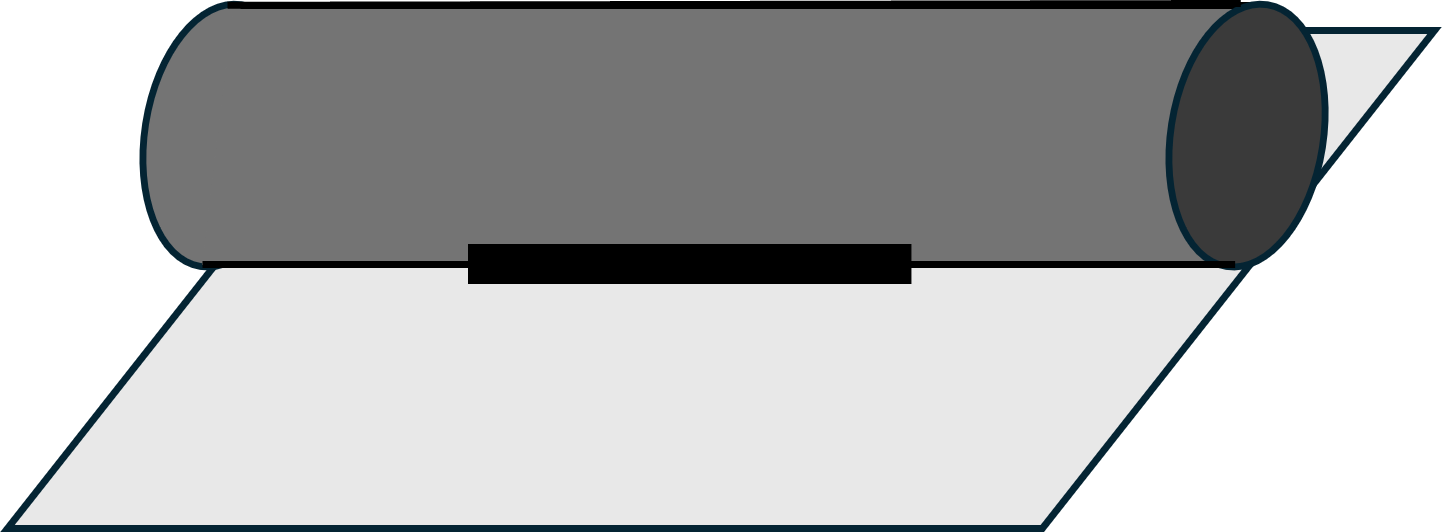}
\caption{\label{catterpillarfig} The Riemann surface we associate with the Riemann-Hilbert problem. The heavy dark line denotes the position of the non-trivial jump matrix while the jump matrix in the rest of the real axis is $\sigma_x$}
\end{figure}

  This equation has the form of a Riemann-Hilbert problem, but unlike more conventional Riemann-Hilbert problems, though, this equation, namely Eq.  (\ref{preRH}), is non-local, requiring the knowledge of $R(x+\imath c,w)$, which does not lie on the real axis, to obtain the jump discontinuity of $R$ on the real axis. To circumvent this problem, we fold the strip $0\leq\Im(z)\leq c$ into a cylinder and glue it onto the real axis. We thus have a  surface with two sheets the lower sheet is a plane. We assign to the lower half plane of the lower sheet the function $R(z,w)$. Namely, we write $R^{(1)}(z,w)=R(z,w)$ for $Im(z)<0$. The superscript $1$ denotes that this it the lower sheet. We assign to the upper half plane of the lower sheet the function $R(z+\imath c,w)$ namely $R^{(1)}(z,w)=R(z+\imath c,w)$ for $\Im(z)>0$. The upper sheet, sheet $2$, is a cylinder and we assign $R^{(2)}(z,w)=R(z,w)$ for $0\leq \Im(z)\leq \imath c$. To give the second sheet the topology of the cylinder, we require $R^{(2)}(z+\imath c,w)=R^{(2)}(z ,w).  $ The surface in question is depicted  in Fig. \ref{catterpillarfig}

Denoting the support of the Bethe roots in the large $N$ limit by the union of intervals,  $\mathcal{S}$, we have the following Riemann Hilbert problem on this newly defined surface:
\begin{align}\label{RH1}
&\begin{pmatrix}R^{(1)}_- &
R^{(2)}_- \\
\end{pmatrix}=
\begin{pmatrix}
R^{(1)}_+ & R^{(2)}_+  \\
\end{pmatrix}\begin{pmatrix}e^{\varphi_-}-e^{\varphi_+} & 1\\1 & 0 \end{pmatrix}, \quad z\in \mathcal{S}\\
&\begin{pmatrix}R^{(1)}_- &
R^{(2)}_- \\
\end{pmatrix}=
\begin{pmatrix}
R^{(1)}_+ & R^{(2)}_+  \\
\end{pmatrix}\begin{pmatrix}0 & 1\\1 & 0 \end{pmatrix}, \quad z\notin \mathcal{S}.\label{RH2}
\end{align}
and $R^{(i)}(z,w)\sim \frac{1}{z}$ at $z \to\infty$ on both sheets. In addition we have $R^{(i)}(z,w)\sim \frac{1}{z-w}$ for $z\to w$ on the appropriate sheet, depending on which sheet $w $ falls on by the convention set above.

\begin{figure}[b]
\includegraphics[width=13cm]{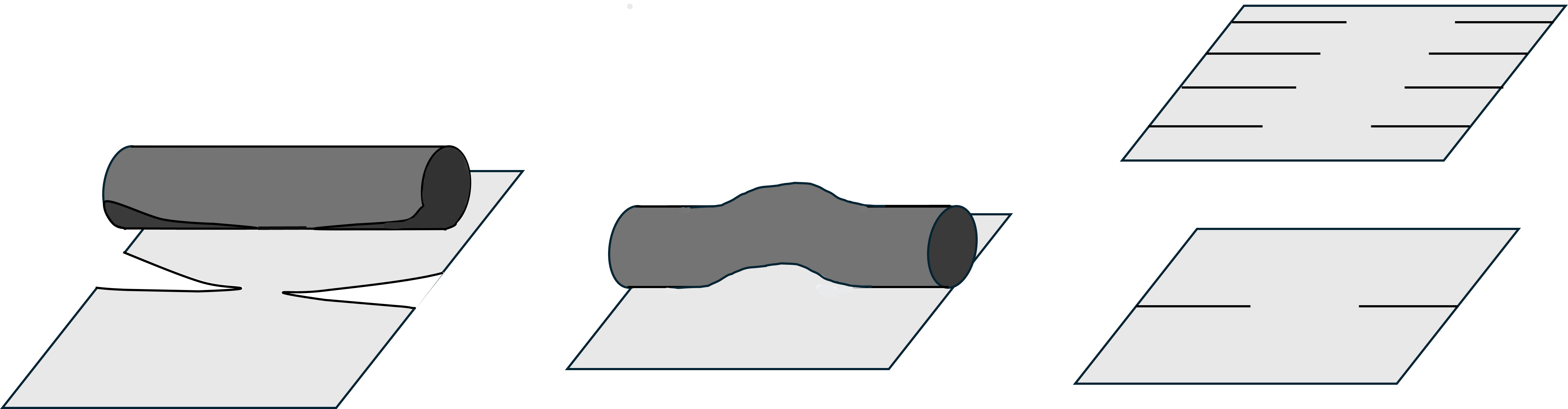}
\caption{\label{cuttingpasting} On  the left a cylinder and a plane are cut along a line. In the middle the result of gluing is shown. The surfaces are glued along the opposing edge of the surface, namely the upper edge of the cylinder is glued to the lower edge of the plane and vice versa. On the right a model of the surface is shown in which the surface is reproduced by taking the usual Riemann surface associated with $\sqrt{z^2-N^2}$ (albeit with unconventional choice of letting the branch cut run through infinity) and reproducing all singularities, including the jump discontinuity produced by the branch cut, periodically with period $\imath c$.  }
\end{figure}

\subsection{The steepest descent method}
After presenting the problem as a Riemann-Hilbert problem on an unconventional surface, the question arises as to what extent is this Riemann-Hilbert problem tractable? A rather immediate course of attack to solve the problem is to apply the steepest descent method of Ref. \cite{Deift:Zhou:Steepest:Descent}. We shall show below that the method is applicable even with the special surface of Fig. \ref{catterpillarfig}. We shall need in fact to endow this surface with the structure of a Riemann surface. This will be discussed in the next sub-section, while in the present sub-section we shall briefly outline the steepest descent method, or rather only parts of it which will be necessary for our purposes.   

Let us define  $S_\pm^{(i)}=\left(\frac{z-N}{z+\\ N}\right)^{\frac{2 i-1}{4}}R_\pm^{(i)}e^{g_\pm^{(i)}}.$ The new function $S$ satisfies a jump condition with a new jump matrix defined as:
\begin{align}
(S_-^{(1)},S_-^{(2)})=(S_+^{(1)},S_+^{(2)})\begin{pmatrix}e^{g^{(1)}_-(z)-g^{(1)}_+(z)}(e^{\varphi_-(z)}-e^{\varphi_+(z)}) & e^{g^{(2)}_-(z)-g^{(1)}_+(z)} \\
-e^{g_-^{(1)}(z)-g^{(2)}_+(z)} & 0 \\
\end{pmatrix}
\end{align}
We let 
\begin{align}
&g^{(1)}_+(z)-g^{(1)}_-(z)=\log\left(e^{\varphi_-(z)}-e^{\varphi_+(z)}\right) \\
&g^{(2)}_-(z)-g^{(1)}_+(z)+g_-^{(1)}(z)-g^{(2)}_+(z)=0
\end{align} which is solved by as follows:
\begin{align}
g(x)=\int dx\int _{x\in\mathcal{S}}d\omega(x)\log\left(e^{\varphi_+(x)}-e^{\varphi_-x)}\right),
\end{align}
where $\mathcal{S}$ is the support of the Bethe roots and $d\omega(x)$ is a differential having a pole singularity on both at point $x$ the residue being $\pm1$ on respective sheets. We relegate the  question of how to write down $d\omega(x)$ on the surface of Fig. \ref{catterpillarfig}. to the next sub-section and proceed for now assuming that this differential is readily available.

After employing the function $g$, just obtained, the equation for $S$ becomes simpler. Indeed, the jump condition  reads:
\begin{align}
(S_-^{(1)},S_-^{(2)})=(S_+^{(1)},S_+^{(2)})\begin{pmatrix}1 & e^{g^{(2)}_--g^{(1)}_+} \\
-e^{-g^{(2)}_-+g^{(1)}_+} & 0 \\
\end{pmatrix}
\end{align}

\begin{figure}[b]
\includegraphics[width=5cm]{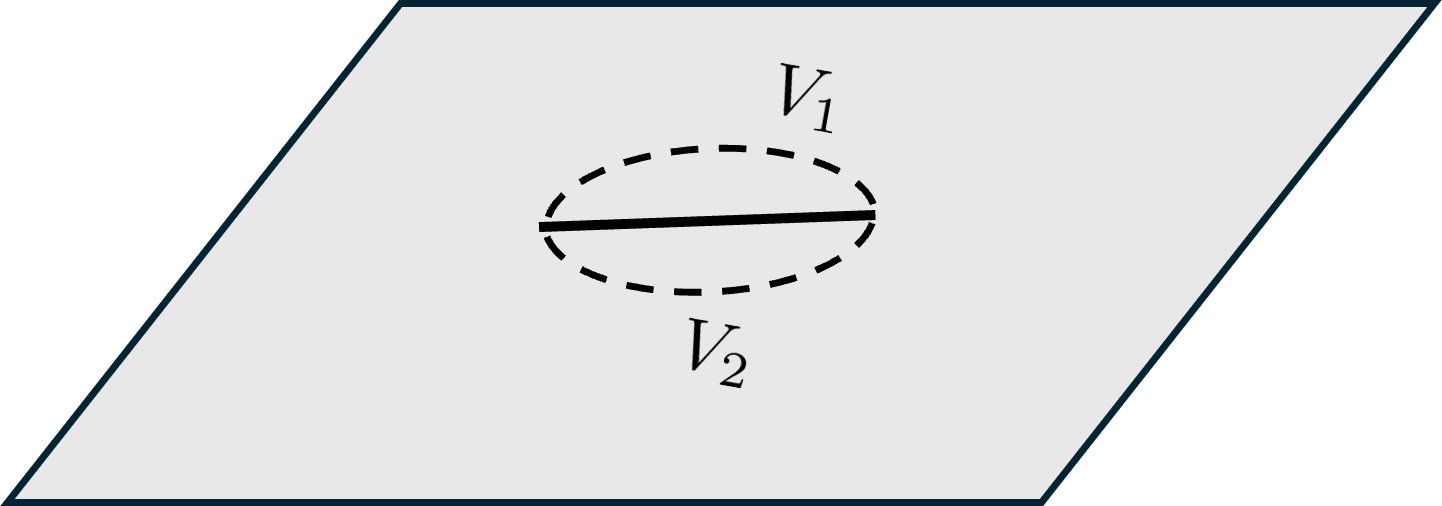}
\caption{\label{lensefig} The original contour on which the jump matrix was defined, the heavy line, which coincides with the support of the Bethe roots, is deformed into two contours (here in dashed lines) each with its own jump matrix, namely $V_1$\ and $V_2$. On the new contours these matrices are the identity to exponential accuracy. The Riemann-Hilbert problem thus is simplified into piecewise constant matrices, which are the identity and $\sigma_x$ in this case. At the two edges of the lens the problem is non-trivial but may be solved separately employing other methods.     }
\end{figure}

We can now decompose the jump matrix above as:
\begin{align}
&\begin{pmatrix}1 & e^{g^{(2)}_--g^{(1)}_+} \\
-e^{g^{(1)}_+-g^{(2)}_-} & 0 \\
\end{pmatrix} =V_1V_2,\\& \quad V_1=\begin{pmatrix}1 & 0 \\
-e^{g^{(1)}_+-g^{(2)}_-} & 1 \\
\end{pmatrix}, \quad V_2=\begin{pmatrix}1 & e^{g^{(2)}_--g^{(1)}_+}  \\
0 & 1 \\
\end{pmatrix} 
\end{align}
The jump contour is separated into two contours, one on which $V_1$ constitutes the jump condition while on the other contour $V_2$ is the appropriate jump matrix. Those contours are then deformed to the upper and lower half planes respectively forming the lenses of Fig. \ref{lensefig}. The exponential, off-diagonal terms in both $V_1$ and $V_2$ quickly become zero, and as such the jump condition becomes trivial. 

To see this more explicitly, one define a new vector, $T$, a vector that will be shown to have  diagonal jump matrix which is the identity outside the support of the Bethe roots and equal to 
the anti-diagonal matrix $\sigma_x$ on the support. One sets  $T\equiv
S$ outside the lens of Fig. \ref{lensefig} on the upper half plane $T\equiv
S V_1$ inside the lens on the upper half plane  then $T \equiv SV_2^{-1}$ inside the lens in the lower half plane an finally $T\equiv S$ in the lower half plane outside the lens.  The solution thus defined, which we call $T$, has the identity as a jump matrix on the support of the Bethe roots and $\sigma_x$ outside. Any meromorphic function on the Riemann surface that will be defined in the next sub-section, will satisfy this jump condition. We proceed then by discussing this Riemann-surface.  

\subsection{The Riemann surface}
We have seen above that that we have to write down a differential $d\omega(x)$ on the surface of Fig. \ref{catterpillarfig} along with  a meromorphic function $T,$ defined on the same Riemann surface. We must endow the surface shown in Fig. \ref{catterpillarfig} with the structure of a Riemann surface. To do so, first ignore the periodicity of the second sheet. Namely, take the second sheet to be a plane rather than a cylinder. Then, $d\omega(x)=\frac{1}{z-x}\sqrt{\frac{w^2-N^2}{z^2-N^2}}dz$ is a differential on that Riemann surface  having the correct behavior, namely a pole of residue $\pm1$ on the respective sheets. We draw the branch cut of $\sqrt{z^2-N^2}$ as to go from $N$\ to infinity through the positive real axis and then return to $-N$ coming from infinity through the negative real axis. A function or a differential on this Riemann surface will then have automatically a $\sigma_x$ jump matrix on the cut,  the matrix $\sigma_x$ simply corresponding to the change of sheets through the cut. 

To recover the periodicity on the second sheet, we then replicate any singularity on the second sheet, periodically with period $\imath c$. 
 This is depicted schematically in Fig. \ref{cuttingpasting}. More concretely, the differential $d\omega(x_0)$ takes the form: 
\begin{align}\label{domega}
&\frac{d\omega(x_0)}{dz} =\frac{1}{z-x_0}\frac{\sqrt{x_0^2-N^2}}{\sqrt{z^2-N^2} }-\frac{1}{2}\sum_{n\neq 0} \frac{1}{z-x_0-n\imath c}\left(1-\frac{\sqrt{(x_0+n\imath c)^2-N^2}}{\sqrt{z^2-N^2}}\right)+\\&+\frac{1}{2}\int_{|x|>N}\sum_{n\neq 0} \frac{1}{z-x-n\imath c} \left(1-\frac{\sqrt{(x+n\imath c)^2-N^2}}{\sqrt{z^2-N^2}}\right)    \rho_{x_0}(x)dx.\nonumber
\end{align}
The first term on the right hand side introduces the poles at $x_0$ on both sheets. Then the second term enforces the periodicity of the pole on the second sheet by replicating this pole only on the second sheet periodically at intervals $\imath c$. The second sheet, then, being looked at as a plane, has periodic poles with equal residues and a single branch cut only on the real axis. The cut is a jump singularity of  $d\omega(x_0),$  when the differential is viewed as a differential on the plane which is the second sheet, rather than a differential on the Riemann surface composed of both sheets. This jump discontinuity must be replicated, which is done by the third term on the right hand side. The jump discontinuity   $2\pi\imath \rho_{x_0}(x)$  must be determined self-consistently, as every term in the sum introduces itself a jump discontinuity on the cut in addition to the jump discontinuity on the cut introduced by the first two terms on the right hand side.    

We now write the equation determining $\rho_{x_0}.$ Only after doing this is the expression for $d\omega(x_0)$\ complete.   First consider that  the analyticity of the differential on the Riemann surface (except at the singularity) requires $d\omega^{(1)}(x+\imath0^+)=d\omega^{(2)}(x-\imath0^+)$
and $d\omega^{(1)}(x-\imath0^+)=d\omega^{(2)}(x+\imath0^+)$. This in turn suggests that the jump discontinuity on the second sheet is   $-2\pi\imath \rho_{x_0}(x)$ and that the average value of $d\omega$ across the cut, 
$\frac{d\omega^{(i)}(x+\imath 0^+)+d\omega^{(i)}(x+\imath 0^-)}{2} $ is the same for both sheets. We can write the average value across the cut by knowledge of the singularities on either sheets and equate them to obtain the following equation that determines $\rho_{x_0}$:
\begin{align}
\dashint \frac{\rho_{x_0}(x')dx'}{x-x'}+\frac{1}{x-x_0}=-\frac{\pi}{c}\dashint \rho_{x_0}(x') {\rm  ctgh\frac{\pi}{c}}\left(x-x'\right)dx'-\frac{\pi}{c} {\rm  ctgh\frac{\pi}{c}}\left(x-x_0\right)  
\end{align}
where the left hand side is the average value of $d\omega$ on the first sheet while the right hand side is the average value of  $d\omega$ on the second sheet. We have used the identity $\sum_n\frac{1}{x-n\imath c} =\frac{\pi}{c}^{}{\rm ctgh}(\frac{\pi x}{c})$The equation can be written as follows:
\begin{align}\label{jumpEq}
-\dashint \rho_{x_0}(x')\left(\frac{1}{x-x'}+\frac{\pi}{c} {\rm  ctgh}\frac{\pi}{c}\left(x-x'\right)\right)dx'=\frac{1}{x-x_0}+\frac{\pi}{c} {\rm  ctgh}\frac{\pi}{c}\left(x-x_0\right) 
\end{align}

This equation is a linear equation in $\rho_{x_0}$ with good behavior such that it may be solved numerically. For example in the limit $c\to\infty$ one recognizes the usual equation for the discontinuity of a meromorphic differential on the standard Riemann surface  $\sqrt{z^2-N^2}$. Indeed in this limit $\frac{\pi}{c}{\rm cthg}\frac{\pi}{c} x\sim \frac{1}{x}$, and the equation has the usual electrostatic interpretation and an easily obtainable solution. 

After obtaining $d\omega(x_0)$, one may use this differential in Eq. (\ref{domega}) supplemented with Eq. (\ref{jumpEq}) in order to apply the steepest descent method to the Riemann-Hilbert problem of Eqs. (\ref{RH1},\ref{RH2}).
We have only discussed certain aspects of the steepest descent method, these may require other differentials than the $d\omega(x_0)$. Nevertheless all these differentials are strongly related, as any of the differentials involved may be thought as multi-poles at finite points or at infinity, where those, in turn, can be composed of dipoles. Since $d\omega(x_0)$ is composed of a pair of poles, it is rather a straightforward extension to write any of the needed differentials, starting from the example of $d\omega(x_0)$. We  thus omit a more general treatment for the sake of brevity. We instead work out below a particular example, where a Slavnov overlap is computed using only the elements discussed above. 

We also need to find $T$ a meromorphic function on the Riemann surface. More precisely $T^{(1)}$ is the value of a meromorphic function on the first sheet and $T^{(2)}$ on the second. This function must have  a pole at $w$\ of residue $1$. This requirement comes from examining the defining equation for $R$, namely Eq. (\ref{ToSolve}). In fact, the function $R(z,w)$ must be eventually integrated in a contour integral around the Bethe roots. This contour can be drawn slightly above the real axis on the original complex plane and integrate over $(R(z,x+\imath 0^+)-R(z,x-\imath0^+))(\dots)dx$, where the ellipsis denotes any other term in the integral. The combination $R(z,x+\imath 0^+)-R(z,x-\imath0^+)$ has two poles with resides given by $\pm1.$ The poles are on opposite Riemann sheets on the surface defined above. This means that we may choose $T$ to be a function with this pole structure, namely as $\frac{d\omega(x)}{dz}$.

\section{Large $c$ limit}

After giving the general method (or rather some essential aspects  of it) for solving Eq. (\ref{ToSolve}), and thus for eventually computing the Slavnov determinant. We shall now demonstrate how the method works for the particular problem of the Anderson orthogonality catastrophe. This problem involves free fermions, so that we take $c\to\infty$ in this section. The problem involves two states, one describing free fermions on a ring with no flux piercing the ring. Namely the Bethe roots are at the point $\frac{2\pi n}{L},$ being just momenta of free particles. This state is take to be the ground state, such that we take the filled Fermi sea given by $-N\leq n\leq N.$ This can be regarded as the on-shell Bethe state in the Slavnov overlap. The other states has a flux piercing the ring, leading to a  change in the quantization condition, giving the Bethe roots  $\frac{2\pi (n+\delta)}{L}$ again at the ground state,  $-N\leq n\leq N$. This is the off-shell state (the roles of on- and off-shell states may be interchanged since a ``twist'', namely the flux condition, may be incorporated into the Bethe equations turn the off-shell state into the on-shell one and vice versa).
The overlap between the two states behaves as $N^{-\delta^2}$, a result due to Anderson\cite{Anderson:Catastrophe}. 
\subsection{Large $c$ limits of Riemann-Hilbert data} 
To set up the problem more concretely, let us assume that we have two states one shifted by $\delta$ with respect to one another. This two sets of Bethe roots are incorporated into one set $\{u_i\}_{i=1}^{4N}$, as this set determines the matrix $K$\ appearing in the Matsuo-Kostov representation of the Slavnov determinant. We thus have the set: 
\begin{align}
u_j=\begin{cases}j-\frac{1}{2} - N & j \leq 2N \\
j-\frac{N+1}{2} - N +\delta & j>2N \\
\end{cases}. 
\end{align}

The function $\varphi(z),$ defined in Eq. (\ref{varphidef}), and which features heavily in the formalism may be computed next:\begin{align}
&\varphi(z)=   -2\pi\imath z + \sum_l \log
  {z-u_l+ \imath c    \over z-u_l}.
\end{align}
This expression can be written through Gamma functions:
\begin{align}  
&\varphi(z)=-2\pi\imath z+\log\frac{\Gamma\left(z-N\right)\Gamma\left(z-\delta-N\right)\Gamma\left(z+\imath c+N\right)\Gamma\left(z-\delta+\imath c+N\right)}{\Gamma\left(z+N\right)\Gamma\left(z-\delta+N\right)\Gamma\left(z+\imath c-N\right)\Gamma\left(z-\delta+N\right)} 
\end{align} 
Using the asymptotics of the $\Gamma$ function,
\begin{align}
\log\Gamma(n)=(n-1/2)\log(n)-1/2\log(2\pi)+\dots
\end{align}
 we can write down the asymptotics of $\varphi$:
\begin{align}
&\label{phiGeneral} \frac{\varphi(z)}{2}=-\pi\imath z+\left(z-N -\frac{\delta+1}{2} \right)\log\left(\lambda-N\right)-\left(z+N -\frac{\delta+1}{2} \right)\log\left(z+N\right)-\\&\nonumber-\left(z+\imath c-N -\frac{\delta+1}{2} \right)\log\left(z+\imath c-N\right)+\left(z+\imath c+N -\frac{\delta+1}{2} \right)\log\left(z+\imath c+N\right)
\end{align}

\subsection{Outer Region}

We now proceed to solve the Riemann-Hilbert problem in the region outside the immediate vicinity of the points $\pm N$, which constitute the endpoints of the support of the Bethe roots. This `outer' region supports a solution that can be obtained by the method outlined above and summarized in Fig. \ref{lensefig}, which involves redefining the Riemann-Hilbert problem by multiplying by the exponent of the $g$\ function.  In this vein, let us define as before $S_\pm^{(i)}=\left(\frac{z-N}{z+\\ N}\right)^{\frac{1}{4}}R_\pm^{(i)}e^{g_\pm^{(i)}}.$ The vector $S$ satisfies the the following jump condition:
\begin{align}
(S_-^{(1)},S_-^{(2)})=(S_+^{(1)},S_+^{(2)})\begin{pmatrix}e^{\varphi_+(z)+g^{(1)}_-(z)-g^{(1)}_+(z)}(e^{-2 \imath z}-1) & e^{g^{(2)}_-(z)-g^{(1)}_+(z)} \\
-e^{g_-^{(1)}(z)-g^{(2)}_+(z)} & 0 \\
\end{pmatrix}
\end{align}
We let 
\begin{align}
&g^{(1)}_+(z)-g^{(1)}_-(z)=\varphi_+(z)+\imath \pi \\
&g^{(2)}_-(z)-g^{(1)}_+(z)+g_-^{(1)}(z)-g^{(2)}_+(z)=0
\end{align} which is solved by as follows. First we write an equation for $\varphi$ in the upper half plane near the real axis, which is obtained from Eq. (\ref{phiGeneral}) by assuming $c\gg N$:
\begin{align}
\frac{\varphi_+(z)}{2}=\left(z-N -\frac{\delta+1}{2}\right)\log\frac{N -z}{\imath c e} -\left(z+N -\frac{\delta+1}{2}\right)\log\frac{N +z}{\imath c e} +2N
\end{align}
Note that the term $-2\pi\imath z$ in Eq. (\ref{phiGeneral}) was incorporated into the logarithmic function by using $z\log\left(\frac{N -z}{\imath c e}\right)  $ instead of  $z\log\left(\frac{z -N}{\imath c e}\right)-\pi\imath z.  $ Then we write a solution to the equations:
\begin{align}
&g_+^{(1)}(z)=+\frac{\varphi_+(z)+\imath \pi}{2}   +\imath \pi z,\quad g_-^{(1)}(x)=-\frac{\varphi_+(z)+\imath \pi}{2}+\imath \pi z,\\
&g_+^{(2)}(x)=-\frac{\varphi_+(z)+\imath \pi}{2}-\imath \pi z,\quad g_-^{(2)}(x)=+\frac{\varphi_+(z)+\imath \pi}{2}-\imath \pi z.
\end{align}
The part of this result which is $O(N),$ namely excluding the term $\frac{\log\frac{z-N}{z+N}}{4}\pm\frac{\imath\pi}{2} ,$  has a simple form on the original plane 
\begin{align}
&2 g(z)=-\left(z-N\right)\log\frac{z-N }{ e} +\left(z+N\right)\log\frac{z+N }{ e} +\\
&+\left(z+\imath c-N\right)\log\frac{z-N }{ e} -\left(z+\imath c+N\right)\log\frac{z+N }{ e} +O(1) 
\end{align}
Then associating with the region around the origin the new $g^{(2)}_+$ and  $g^{(1)}_-$ and the region around $\imath c$ with $g^{(1)}_+$ and  $g^{(2)}_-$ with $z\to z-\imath c$, we get the result above. 

Note that the function $g$ has a logarithmic branch cut from $-N$ to $-N+\imath  c.$ However, the jump across the branch cut is an integer multiple of $2\pi\imath$ and thus does not affect $e^g$ which is smooth on this cut.

After employing the function $g$, just obtained, the equation for $S$ becomes simpler in the outer region. Indeed, the jump condition outside a small vicinity around the edges reads:
\begin{align}
(S_-^{(1)},S_-^{(2)})=(S_+^{(1)},S_+^{(2)})\begin{pmatrix}1 & e^{g^{(2)}_--g^{(1)}_+} \\
-e^{-g^{(2)}_-+g^{(1)}_+} & 0 \\
\end{pmatrix}
\end{align}
Where  we  have bent the contour to the lower half plane so we could replace $1-e^{-2\imath z}$ with $1$. We must only deform the contour an amount which is of order $1$, namely much smaller than the size of the contour itself, which is of order $N$. We can decompose the jump matrix above as:
\begin{align}
&\begin{pmatrix}1 & e^{g^{(2)}_--g^{(1)}_+} \\
-e^{g^{(1)}_+-g^{(2)}_-} & 0 \\
\end{pmatrix} =V_1V_2,\\& \quad V_1=\begin{pmatrix}1 & 0 \\
-e^{g^{(1)}_+-g^{(2)}_-} & 1 \\
\end{pmatrix}, \quad V_2=\begin{pmatrix}1 & e^{g^{(2)}_--g^{(1)}_+}  \\
0 & 1 \\
\end{pmatrix}, 
\end{align}
whereupon expanding to lenses creates a new, trivial problem. One sets  $T=S$ outside the lens on the upper half plane $T=S V_1$ inside the lens on the upper half plane  then $T =SV_2^{-1}$ inside the lens in the lower half plane an finally $T=S$ in the lower half plane outside the lens.  The solution thus defined, which we call $T$, has the identity as the jump matrix outside the support of the Bethe roots and a jump matrix equal to 
the anti-diagonal matrix $\sigma_x$ on the support.

Then inside the lens in the upper half plane we have
\begin{align}
&\nonumber (R(z+\imath c, w)\quad R(z,w))=\left(e^{-g_+^{(1)} }T^{(1)}_++e^{-g^{(2)}_-}T^{(2)}_+\quad -e^{-g^{(2)}_+}T_+^{(2)}\right)  
\end{align}
and inside the lenses in the lower we have
\begin{align}
&\begin{pmatrix}R(z,w) &
R(z+\imath c,w) \\
\end{pmatrix}=\left(e^{-g_-^{(1)}}T^{(1)}_-\quad e^{-g^{(1)}_+}T^{(1)}_-+e^{-g_-^{(2)}}T^{(2)}_-\right)
\end{align}
Outside the lenses $R=Te^{-g}$. 
Looking at the relative magnitude of the different terms, one sees that several elements of $T$ are exponentially small with respect to other, where the exponent is of order $N$ (which is inherited from $|z|= O(N)$). This allows us to write:
\begin{align}
&\label{Ronlyimportant}R(z,w)=e^{+\frac{\varphi_+(z)}{2}+\imath \pi z}\begin{cases}O(1) & \Im(z)>0 \\
O\left(e^{-N}\right) & \Im(z)<0 \\
\end{cases}\\
&\label{R+iconlyimportant}R(z+\imath c,w)=e^{-\frac{\varphi_+(z)}{2}+\imath \pi z}\begin{cases}O\left(e^{-N}\right) & \Im(z)>0 \\
O\left(1\right) & \Im(z)<0 \\
\end{cases}
\end{align}
Such that $R$ actually is $0$ to exponential accuracy below and above the cut around the support of the Bethe roots and the support of the Bethe roots shifted by $\imath c$, respectively.

At this moment $T$ is chosen in the following form 
\begin{align}
T(z,w)=\frac{1}{2}\frac{1}{z-w} \left(1+\frac{\sqrt{w^2-N2}}{\sqrt{z^2-N^2}}\right)
\end{align}
Then we can write the approximation that we have attained for $R(z,w)$ and $R(z+\imath c,w)$, written now no the original surface, namely without the conventions associated with the sheets of the Riemann surface adopted before for the branches of the different roots:
\begin{align}
\label{Routside}R(z,w)&=\frac{1 }{2}\frac{1}{z-w} \left(1+\frac{\sqrt{w^2-N^2}}{\sqrt{z^2-N^2}}\right)\sqrt[4]{\frac{z-N}{z+N}\frac{w+N}{w-N}}e^{g(z)-g(w)}\\
\label{R+icoutside}R(z+\imath c ,w)&= \frac{1}{2}\frac{1}{z-w} \left(1-\frac{\sqrt{w^2-N^2}}{\sqrt{z^2-N^2}}\right)\sqrt[4]{\frac{z-N}{z+N}\frac{w+N}{w-N}}e^{-g(w)-g(z)}.
\end{align}

\subsection{Inner Region}

The solution just obtained for $R(z,w)$ is valid away from small vicinity of the edges of the support of the Bethe roots. This solution is sufficient to derive the   leading order of the logarithm of the overlap, which is typically of order $N$. Nevertheless, here the result for the logarithm of the overlap is known to be $-\delta^2\log(N)$ \cite{Anderson:Catastrophe}. Therefore, an otherwise sub-leading order is required. To find this ostensibly sub-leading order it is necessary to solve the problem around the endpoints of the support of the Bethe roots to better accuracy. This solution is to be found by demanding that this inner solution matches the solution away from the endpoints, namely the exterior solution.   

We propose solutions in the  inner region next to the points $\pm N$.  We posit the following solution for $z$ around $\imath c$, namely in the vicinity of the Bethe roots after a shift of $\imath c$:
\begin{align}
R(z+\imath c,w)=r^{+\imath c}(z,w)\frac{\Gamma\left(  z+N+1- \alpha\right)}{(\imath c)^{2N}\Gamma\left(  z-N- \alpha\right)},
\end{align} 
where $r^{+\imath c}$ is a smooth function.  Taking into account the equation for $R, $ Eq, (\ref{ToSolve}) we have the behavior of $R$ around the origin, namely in the vicinity of the Bethe roots:
\begin{align}
& R(z,w)=    \frac{(\imath c)^{2N}r^{+\imath c}(z,w)e^{-\imath 2\pi  z}\Gamma\left(  z-N\right)\Gamma\left(  z-N- \delta\right)\Gamma\left(  z+N- \alpha\right)}{\Gamma\left(N+1+  z\right)\Gamma\left(  z+N+1- \delta\right)\Gamma\left(  z+N+1- \delta\right)\Gamma\left(  z-N-  \alpha\right)}+\dots
\end{align}   
where the ellipsis denotes terms which are analytic in this region. We divide  this by the factor we encounter in the exterior region which is $\frac{\Gamma\left(  z-N- \beta\right)}{\Gamma\left(  z+N+1- \beta\right)}$ to define $r(z,w)$:
\begin{align}
&r(z,w)\equiv \frac{\Gamma\left(  z+N- \beta\right)}{(\imath c)^{2N}\Gamma\left(  z-N- \beta\right)} R(z,w)+\dots\label{rtoR}
\end{align} 
We can analyze the behavior of the residues of this term by first re-writing it, using $\Gamma(  x)\Gamma(1-  x)=\frac{\pi}{\sin(\pi   x)}$, as follows:
\begin{align}
& r(z,w)=r(z,w;\delta,0)\quad\mbox{where}\\
&\label{rLotsofGammas}r(z,w;\delta_1,\delta_2)=\frac{e^{-2\imath \pi z}r^{+\imath c}(z,w)\sin(\pi(  z-  \alpha))\sin(\pi(  z-  \beta))}{\sin(\pi  (z+\delta_2))\sin(\pi(  z-  \delta_1))}\times\\&\times\nonumber\frac{\Gamma\left(N+  z- \beta\right)\Gamma\left(N+1-  z+ \beta\right)\Gamma\left(N+  z- \alpha\right)\Gamma\left(N+1-  z+  \alpha\right)}{\Gamma\left(N+1+  z+\delta_2\right)\Gamma\left(N+1-  z-\delta_2\right)\Gamma\left(N+1-  z+ \delta_1\right)\Gamma\left(N+1+  z-\delta_1\right)}+\dots
\end{align}
Here we have added the variables $\delta_1,$ and $\delta_2$ for later convenience. At this moment we let $\delta_1=\delta$ and $\delta_2=0.$ Later, when relevant, we shall first take a derivative with respect to $\delta_1$ and $\delta_2$ and only then set them equal to $\delta$ and $0,$ respectively.  

The poles of this expression, their location denoted by $a_n^{(1)},$ $a_n^{(2)}$ with  $ N\leq n\leq -N $, are given by  $a_n^{(1)}=n$ and   $a_n^{(2)}=n+\delta.$ Defining $\delta n=N+1-n$
and $\tilde \delta n=N+1+n,$ given the following choice for   $  \alpha $ and  $  \beta$,\begin{align}
 \alpha= \beta=\frac{ \delta}{2}-\frac{1}{4},\label{choice1}
\end{align}
are given by:
\begin{align}
&\underset{  z\to a_n^{(i)}}\Res r(z,w)=\pm r^{+\imath c}( a_n^{(i)},w)\frac{\sin^2\pi\left(\frac{  \delta}{2}\mp\frac{1}{4}\right)}{\sin(\pi  \delta)}\times\frac{e^{-\imath \pi \delta(1\mp1)}}{\tilde \delta n^{\frac{1}{2}}\delta n^{\frac{1}{2}}},
\end{align}
where the signs are to be taken respectively to $i=1,2$.  

Taking into account both sets of poles, $a_n^{(1)},$ $a_n^{(2)}$ and with the following choice for $r^{+\imath c}$
 \begin{align}
&r^{+\imath c}(z,w)=\frac{e^{\imath \pi \delta}}{\imath-\cos(\pi  \delta)}\frac{\sqrt{w^2-N^2}}{z-w}\sqrt[4]{\frac{w+N}{w-N}},
\end{align}
one obtains 
\begin{align}
r(z,w)=\sqrt[4]{\frac{w+N}{w-N}}\frac{\sqrt{w^2-N^2}}{z-w}+\dots
\end{align}
with the ellipsis denoting terms which do not jump over the real axis as a function of $z$.  Then, combining this with Eq. (\ref{rtoR}), one obtains that the choice of $\alpha$ and $\beta$ above matches the  form of $R(z,w)$ and $R(z+\imath c,w)$ in the exterior region, Eq. (\ref{Routside}-\ref{R+icoutside}), above and below the the support of the Bethe roots, respectively. The latter condition, whereby the matching of the asymptote being done only at these points, being due to the exponentially small values at the locations that are not matched according to Eqs. (\ref{Ronlyimportant}-\ref{R+iconlyimportant}).

In order to determine $F$\ from Eq. (\ref{OurF}), we now multiply by $\sum_{j}\frac{1}{u_i-u_j}$. This means multiplying by $\frac{L}{2\pi}\sum_{m=-n}^\infty\frac{\mp1}{a_m^{(i)}- a_m^{(3-i)}}$, thus

\begin{align}
& \underset{  z\to a_n^{(i)}}\Res r(z,w) \sum_{m}\frac{\mp1}{ a_n^{(i)}- a_m^{(3-i)}}=-\frac{d}{d \delta_i}\left. \underset{  z\to a_n^{(i)}}\Res r(z,w; \delta_1,\delta_2)\right|_{ \delta_1=\delta, \delta_2=0}=\\&=\left( \frac{ \delta}{\delta n}+\frac{ \delta}{\tilde \delta n}+\cot(\pi \delta)\pm\log\frac{\delta n}{\tilde \delta n}\mp\frac{1}{2}\frac{1}{\delta n}\pm\frac{1}{2}\frac{1}{\tilde \delta n}\dots\right)\underset{  z\to a_n^{(i)}}\Res r(z,w)
\end{align}
This result is obtained by applying a derivative with respect to $ \delta_1 $ and $\delta_2$ of  Eq. (\ref{rLotsofGammas}), while making use of the  expansion of the relevant Gamma functions and the sine function.

We conclude 
\begin{align}
&F^{(i)}(z,w)=R^{(i)}(z,w)\left( \frac{-2N\delta-C(\delta)z }{z^2-N^2}+\cot(\pi  \delta)+C(\delta)\log\frac{z-N}{z+N} \right).
\end{align} 
The constant $C(\delta)$\ does not appear in the final result so we neglect writing an explicit expression for it, which turns out to be rather cumbersome. 

We are now ready to take the integral to determine
the main result of this section by using Eq.(\ref{Fointment}). First we write:
\begin{align}
&\partial_\delta \log\det(\mathds{1}-K)=-2\partial_\delta\sum_{i\in A,j\in A^C }\log(u_i-u_j)+\oint\oint\frac{F(z,w)}{w-z}\frac{dz}{2\pi\imath}\frac{dw}{2\pi\imath}
\end{align} 
We can write the first term on the left hand side as follows:
\begin{align}
&-\partial_\delta\sum_{i\in A,j\in A^C }\log(u_i-u_j)\nonumber=\partial_\delta\log \frac{\prod_{i,j\in A, i<j}(u_i-u_j)\prod_{i,j\in A, i<j}(u_i-u_j)}{\prod_{i\in A, j\in A^C}(u_i-u_j)}=\\& =-\partial_\delta \delta^2 \log(2N)+\dots\label{simpleresult} 
\end{align}
The first equality, which gives an expression for the left hand side using the Cauchy determinant,   is a simple consequence of the fact that the products in the numerator on the right hand side do not depend on $\delta$. The second equality stems from the well-known large $N$ expression for the Cauchy determinant, which is a consequence of the Fisher-Hartwig theorem, given that the Cauchy determinant in this case is also a Toeplitz determinant.
In fact, for this simple case, resorting to the Fisher-Hartwig theorem is not necessary, as one can use basic notion of electrostatic theory in two dimensions, in which context the expression on the right hand side is the electrostatic potential of a set of charges of charges $\pm1$ placed at the point $u_i$, where the sign depends on whether $i\in A$ or $i\in A^C$. 

We can now take the integral of $F(z,w)$ in order to proceed with a calculation of Eq. (\ref{Fointment}):\begin{align}
&\oint\frac{F(z,w)}{w-z}\frac{dz}{2\pi\imath}\nonumber=\left.\partial_z \left( \frac{-2N\delta }{z^2-N^2}+\dots\right)\frac{1}{2}\left(1+\frac{\sqrt{w^2-N^2}}{\sqrt{z^2-N^2}}\right)\sqrt[4]{\frac{z-N}{z+N}\frac{w+N}{w-N}}e^{g(z)-g(w)}\right|_{w=z}=\\
&=-\frac{2N\delta }{w^2-N^2}g'(w)+\dots=\frac{2\delta N}{w^2-N^2}\log\frac{w+N}{w-N}+\dots 
\end{align}  
Where the ellipsis denote all terms that will not contribute or contribute to lower order when the contour integral over $w$ is to be taken. The result of that integration yields:
\begin{align}
\oint \oint \frac{F(z,w)}{w-z}\frac{dz}{2\pi\imath} \frac{dw}{2\pi\imath}=2\delta\log N+\dots
\end{align}

Combining with the result of Eq. (\ref{simpleresult})
we obtain
\begin{align}
\Re(\log\det(\mathds{1}-K))=-\delta^2 \log N+\dots.
\end{align}
This is the well known result for the Anderson orthogonality catastrophe\cite{Anderson:Catastrophe}. 
\section{Conclusion}
In this paper we have presented a method to compute overlaps in quantum integrable models, based on a combination of determinantal formulas for Slavnov overlaps, the functional approach to the computation of the determinant and the Riemann-Hilbert approach to the solution of the functional equations. In order to keep the complexity of the exposition of the method to a minimum, we have chosen to demonstrate the method on the free fermion limit of the Lieb-Liniger model, corresponding to $c\to\infty$, and compared to the Anderson catastrophe result. Indeed, in Refrs. \cite{Imambekov:Glazman:Fermi:Edge:Lieb:Liniger,Cheiakov:Pustilnik:Fermi:Edge:Lieb:Liniger} it was observed that the shift in energy levels at the edge of the support of the Bethe roots dictates Anderson orthogonality exponents and other closely related exponents, for finite $c$. This is to be expected also from the results here, as the this universal behavior seems to originate from the solution of the Riemann Hilbert problem next to the edges of the support. The universality of the results is expected since the leading order of the solution is dictated only by a single parameter, namely the shift of the two sets of Bethe roots near the end points of the support of the Bethe roots.

We stress here that the method shown here can be combined with numerics that solve the Bethe ansatz in the thermodynamic limit. The density of the Bethe roots could be used as input in solving the Riemann-Hilbert problem numerically, as the steepest descent method is amenable to numerical calculation. The fact that numerics is resorted to, would not diminish from the fact that the solution is to be considered an exact solution. Indeed, as is well known, a solution to a many body interacting quantum problem is considered exact, when the complexity of solving the problem does not scale with $N$, the number of degrees of freedom in the problem. Indeed, on the other side of the spectrum of `solutions', lies the brute force approach, namely methods, the computational complexity of which scale exponentially with $N$. 

We should also point out that the methods shown here can be used will little change to  other rational integrable models, such as the $XXX\ $ Heisenberg chain with inhomogeneities, so long as the states involved do not have Bethe strings. To incorporate the latter, an extension of the method must be devised. Though we believe that these extensions may be achieved, the description of how this should be done is beyond the scope of this paper, and is left to future work.    

One  interesting application of such extensions would be to apply this method to  the Kondo model \cite{Andrei:Lowenstein:Review,AndreiL:First:Kondo,Bettelheim:Kondo,Wiegmann:First:Kondo}, the solution of which is closely related to the $XXX$ Heisenberg chain with inhomogeneities. It seems that  similar methods may be applied, but that Bethe strings must play an important role. The expectation is that  the periodicity on the Riemann surface, which in the repulsive Lieb-Liniger model is expressed by the cylinder topology of the upper sheet, may reflect more generally the existence of Bethe strings. It would then be interesting from both the standpoint of the study of the Riemann-Hilbert problem and the study of the Kondo and other models to understand how the Bethe string configuration is reflected in the Riemann-Hilbert problem.

\end{document}